\documentclass{ieeeaccess}
\usepackage{cite}
\usepackage{amsmath,amssymb,amsfonts}
\usepackage{algorithmic}
\usepackage{graphicx}
\usepackage{textcomp}
\usepackage{color,soul}
\def\BibTeX{{\rm B\kern-.05em{\sc i\kern-.025em b}\kern-.08em
    T\kern-.1667em\lower.7ex\hbox{E}\kern-.125emX}}
\begin{document}
\bibliographystyle{IEEEtran}
\history{Received May 14, 2019, accepted June 6, 2019, date of publication June 10, 2019, date of current version July 18, 2019.}
\doi{10.1109/ACCESS.2019.2922113}

\title{ A Design of  Cooperative Overtaking Based on Complex Lane Detection and Collision Risk Estimation}
\author{
\uppercase{JunLan Chen}\authorrefmark{1}, {}
\uppercase{Ke Wang}\authorrefmark{2}, {}
\uppercase{HuanHuan Bao}\authorrefmark{3}, {}
\uppercase{Tao Chen}\authorrefmark{3}, {}
\address[1]{School of Economics \& Management, Chongqing Normal University, Chongqing 401331, China}
\address[2]{State Key Laboratory of Mechanical Transmission, Chongqing University, Chongqing 400044, China}
\address[3]{China Automotive Engineering Research Institute Company, Ltd., Chongqing 401122, China}}
\tfootnote{The authors thank the financial support of National Natural Science Foundation of China (Grant No: 51605054), Key Technical Innovation Projects of Chongqing Artificial Intelligent Technology (Grant No. cstc2017rgzn-zdyfX0039), Chongqing Social Science Planning Project (No:2018QNJJ16), Fundamental Research Funds for the Central Universities (No: 2019CDXYQC003)}
\markboth
{Ke Wang and JunLan Chen \headeretal: A Design of  Cooperative Overtaking Based on Complex Lane Detection and Collision Risk Estimation}
{Ke Wang and JunLan Chen \headeretal: A Design of  Cooperative Overtaking Based on Complex Lane Detection and Collision Risk Estimation}

\corresp{Corresponding author: Ke Wang (e-mail: kewangcqu@163.com);}

\begin{abstract}
Cooperative overtaking is believed to have the capability of improving road safety and traffic efficiency by means of the real-time information exchange between traffic participants, including road infrastructures, nearby vehicles and others. In this paper, we focused on the critical issues of modeling, computation, and analysis of cooperative overtaking and made it playing a key role in the road overtaking area. In detail, for the purpose of extending the awareness of the surrounding environment, the lane markings in front of ego vehicle were detected and modeled with Bezier curve using an onboard camera.  While the nearby vehicle positions were obtained through the vehicle-to-vehicle communication scheme making assure of the accuracy of localization. Then, Gaussian-based conflict potential field was proposed to guarantee the overtaking safety, which can quantitatively estimate the oncoming collision danger. To support the proposed method, many experiments were conducted on the human-in-the-loop simulation platform. The results demonstrated that our proposed method achieves better performance, especially in some unpredictable nature road circumstances.
\end{abstract}


\titlepgskip=-15pt

\maketitle

\section{Introduction}
\label{sec:introduction}
\sethlcolor{yellow}
\PARstart{D}{riving} 
on the road, the safety issues are always the most major concerns, while a huge number of injuries and deaths reveal the story of the global crisis on this topic\cite{RN4727}. Around the world, approximately 1.35 million people die as a result of road traffic crashes \cite{Nobody01} and 50 million are injured per year. While, vehicle overtaking is one of the important causes of casualties\cite{RN5815}. Among them, most of the accidents were caused by the incomprehensive understanding of the nearby environment \cite{RN6986} and the impulsive lane changing behaviors in the traffic stream\cite{RN1429}.

	Overtaking is generally affected by environment understanding and vehicular interactions. Hence, in some framework of intelligent vehicles, sophisticated environment perception module and vehicle-to-vehicle (V2V) wireless communication approach are used, aiming to enable automated cooperation among different vehicles on the road. With the extended capability of situational awareness, cooperative overtaking can enable drivers and virtual-drivers a longer perception range even beyond the field of view. This aspect enables better driving decisions for overtaking. Despite these application advantages, however, there has been comparatively fewer application on this cooperative driving field. Right now, most of the works focused on the communication scheme named VANETs (Vehicular Ad-hoc Networks)\cite{RN4746} discussing the fundamental problem of the communication network via wireless links. It has just solved low-level aspects of ad-hoc networks and standards. But, in the high application level, the practical use of V2V systems in the Advanced Driverless Associate System (ADAS) and Intelligent Vehicle (IV) is rarely noticed. How to promise a high safety benefit for overtaking procedure is still a critical issue and challenging task for intelligent vehicles. 

In this paper, we exploit the cooperative overtaking problem by integrating collision probability model into driving decision procedure. We show that, with the fusion of information from vision sensors and V2V sensors, the risk estimation of overtaking procedure could become more robustness and efficiency. 

 Compared to the state-of-the-art works, there are three main contributions of this paper, which were shown as follows:

(1)  A novel BDI based multi-vehicle collaboration framework was proposed, which uses different kinds of heterogeneous sensors, to extend the awareness of the surrounding environment;

(2) Both the Bezier curve and hybrid Gaussian anisotropic filter were adopted in lane marking detection and modeling algorithm to increase the accurateness and robustness of the estimation of the relative position of the ego vehicle with respect the forward lanes

(3) Gaussian-based probability density function and conflict potentials field were constructed to describe the uncertainty of the driving risk, which can quantitatively estimate the collision probability between nearby vehicles.

The remainder of this paper is organized as follows: the related works were given in Section 2, and the system architecture of the proposed approach was presented in Section 3. Then, the Bezier based lane detection and modeling method were given in Section 4. The collision probability prediction method was discussed in Section 5. At last, the proposed method is validated in Sections 6, and the conclusions are given in Section 7.

\section{ Related Works}
Overtaking is one of the most complex maneuvers for intelligent vehicles both in manual driving mode and automation driving mode\cite{RN4745}. In general, overtaking compose of some consecutive maneuvers. The lane changing behavior followed by traveling a planned path in the adjacent lane parallelled to the overtaken vehicle, and then, a change to the original lane. During the process, different kinds of sensor-based environment perception modules and the longitudinal-lateral motion control modules are comprehensively been undertaken. The two modules are not only associated but also interactive: the environment understanding is the prerequisite, and the following control is the purpose\cite{RN4982},  while both of them are discussed in this paper.

How to use the environment information to make the overtaking procedure safely and smoothly is a key problem\cite{RN1456}. In this end, camera-based driver assistance systems have been equipped in some intelligent vehicles in order to make the front lanes observed by ego vehicle automatically\cite{RN474}. However, because of the regular damage, fracture, and pollution, road marking lines sometimes not clear, the promising detection result cannot be effectively guaranteed, no mention of the noise, light unevenness, water and stains in the real road. Parabola\cite{RN5817}, spline curve\cite{RN5818}, and hyperbola \cite{RN5819} were chosen by researchers to model the road. And then, the lane model parameters are estimated by means of maximum posterior probability estimation\cite{RN5821}. However, due to the time-varying, complexity, nonlinearity, and uncertainty, it is normally hard to obtain accurate mathematical lane models\cite{RN14}. For the purpose of increasing the calculation speed, RANSAC (RANdom SAmple Consensus) algorithm was used in many types of research to eject most of the outliers in the feature matching step \cite{RN5822}.

Meanwhile, in order to estimate the relative positions of nearby vehicles, radar, Lidar, camera, and wireless V2V communication schemes have been explored by some researchers \cite{RN4985}. As the positive kinds of sensors, thanks of their inherent activeness characteristics, radar and Lidar can get the relative distance and velocity directly, however, their performance would drop dramatically in the fog, haze and rain days \cite{RN4984}. For the passive sensors, such as camera, they can only be used in the daytime and lost their capability in the night\cite{RN1499}. Meanwhile, V2V sensors were also comprehensively used in this area, including DSRC (Dedicated Short Range Communications) and IEEE 802.11p protocol \cite{RN5823}. As a wireless short-to-medium-range communications method, DSRC has the capability of permitting high data transmission. While, the IEEE 802.11p/Wireless Access has promulgated a suite of physical and medium access control layer specifications to enable communications in vehicular networks, which enables automated cooperation between different vehicles and road infrastructures \cite{RN5824}. In a unified on-road vehicular network, each connected vehicle periodically transmit and receive position, perception, and safety-related messages with switching on a CCH (common control channel). Use this information, the vehicle can tune into one of the available service channels (SCHs) to exchange all the driving-related information \cite{RN5825}. Compared to radars, Lidar, and camera, the V2V communication scheme can give out a super long range to the vehicle position information. 

In order to plan an overtaking manoeuvre safely, the ego vehicle uses environment perception data and subject vehicle state data to check feasibility of the manoeuvre and design a collision free and safe local reference trajectory for an overtaking manoeuvre \cite{RN6995}. The local trajectory planning can be defined as real time planning of the vehicle’s transition from one feasible state to the next, while satisfying the vehicle’s kinematic limits based on vehicle dynamics and constrained by occupant comfort, lane boundaries and traffic rules, while, at the same time, avoiding obstacles \cite{RN6994}. Meanwhile, there are four well known techniques, including: potential fields, cell de-composition, interdisciplinary methods and optimal control, which were normally employed to construct the local trajectory planning method \cite{RN6992}. 

Although various land marking detection algorithms and plentiful localization strategies as mentioned before have been proposed and applied in the intelligent vehicle area, a combination of the two or the associated application into overtaking risk estimation are really few\cite{RN5826}. Among the previous studies, the overtaking assistant modules usually based on the assumption that the vehicles are going straight with constant speed, which would restrict the performance\cite{RN5828}. Besides, some researchers predefined a safety cell around each vehicle, and warning would be aroused by the disturbance of nearby vehicles\cite{RN5827}. By adopting the safety cell method, the traffic flow rate would reduce seriously. From the works, we can found that, right now, the collision risk in complex traffic is still tough to estimate, which need to be further discussed.

\Figure(topskip=0pt, botskip=0pt, midskip=0pt)[width=320pt]{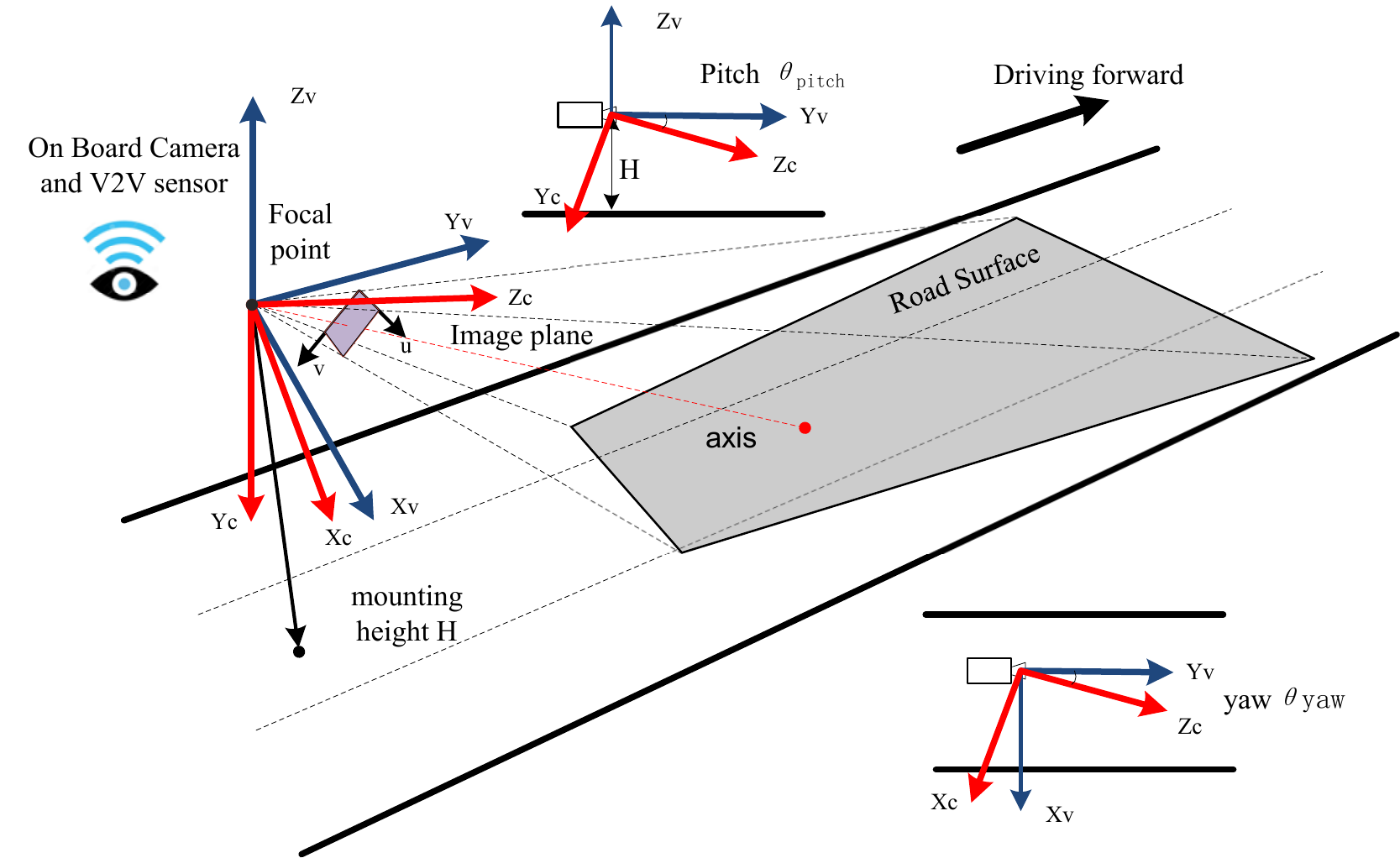}
{Coordinates of rigid body motion model with the forward facing monocular camera and the onboard V2V sensor. \label{fig1}}

\section{Assumption, Coordinates, and collaboration framework}
It is conspicuous that the proposed solution should avert from computationally demanding strategy and aim at developing a method with the ability to operate in real time and accuracy. Therefore, the emphasis of this paper was placed on developing efficient collision risk estimation scheme with the help of the front lane detection and nearby vehicle positions. 
\subsection{Assumption and Coordinates Definition}

Considering a sensor system including a forward facing monocular camera and an onboard V2V sensor, we assume that the ego-motion estimation model is a rigid body motion model and the monocular camera can be modeled by the pinhole camera model and vehicles travel on structured roads, including highways and city roads. Then, using the Zhang method \cite{RN5829}, the intrinsic and extrinsic parameters can be easily calibrated in advance. With the calibration matrix, the relative pose transformation matrix between two sensors can be obtained. Hence, we can use a single coordinate system for both the V2V sensor and the forward facing monocular camera. The coordinate systems are defined in the following (see Figure 1 for illustration).

There are two coordinates used in our system, the global V2V based original global coordinate$\left\{ {{C}_{V}} \right\}$ and the forward facing monocular camera coordinates system$\left\{ {{C}_{C}} \right\}$, which is defined as follows: 
(1) we paralleled ${{X}_{V}}$-${{O}_{V}}$-${{Y}_{V}}$ plane of $\left\{ {{C}_{V}}\right\}$ to the horizontal plane. The ${{Z}_{V}}$-axis points opposite to gravity. The ${{Y}_{V}}$-axis points forward of the vehicle platform, and the ${{X}_{V}}$-axis is determined by the right-hand rule.
(2) forward camera coordinate system$\left\{ {{C}_{C}} \right\}$ is set originated at the optical center of the camera system. The ${{X}_{C}}$-axis points to the left, the ${{Y}_{C}}$-axis points downward, and the ${{Z}_{C}}$-axis points forward coinciding with the camera principal axis.

\subsection{BDI based V2V collaboration framework}
The communication mechanism satisfies the standard of IEEE 802.11p protocol, which was used as the information interaction channel. In this way, the effective communication distance is of 200 meters and the information was packaged by the basic safety message (BSM), including sharing information and behavioral collaboration information. In the context of wireless communication based vehicular network, belief-desire-intention (BDI) framework was adopted to construct a hierarchical and cooperation interaction model, which was shown in Figure 2. 

In the framework, the belief module was mainly used to realize the environment perception task, especially for road environment perception using the onboard camera and V2V sensors. With the belief module, the system can monitor the nearby environment and vehicle states in real time, including lane detection and relative distance between nearby vehicles. While, the desire module was designed for environment risk evaluation and risk estimation, such as the collision estimation, which is the core function of the vehicle cooperation system. Then, the intention module is used to achieving the task of behavioral cooperation, such as path planning, cruise control, and overtaking collaboration. The detail is shown in Figure 3. 

\Figure(topskip=0pt, botskip=0pt, midskip=0pt)[width=380pt]{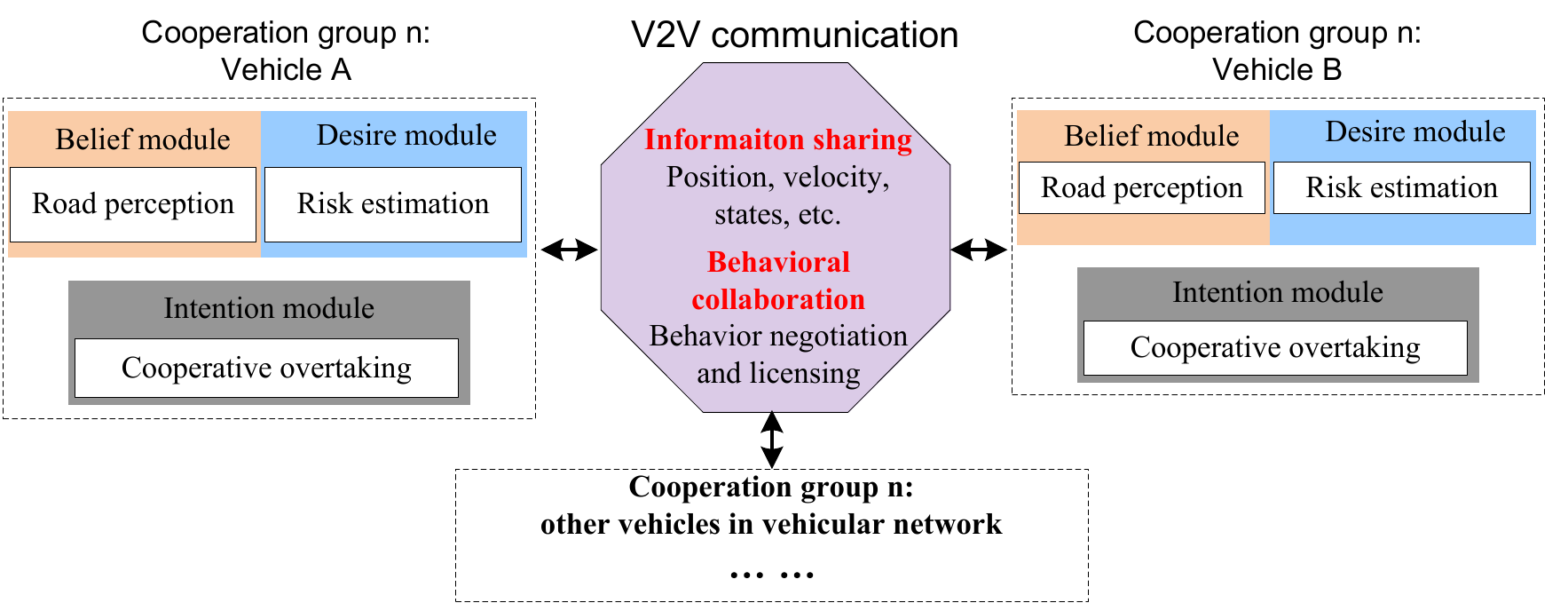}
{BDI based V2V collaboration framework. \label{fig2}}
\Figure(topskip=0pt, botskip=0pt, midskip=0pt)[width=380pt]{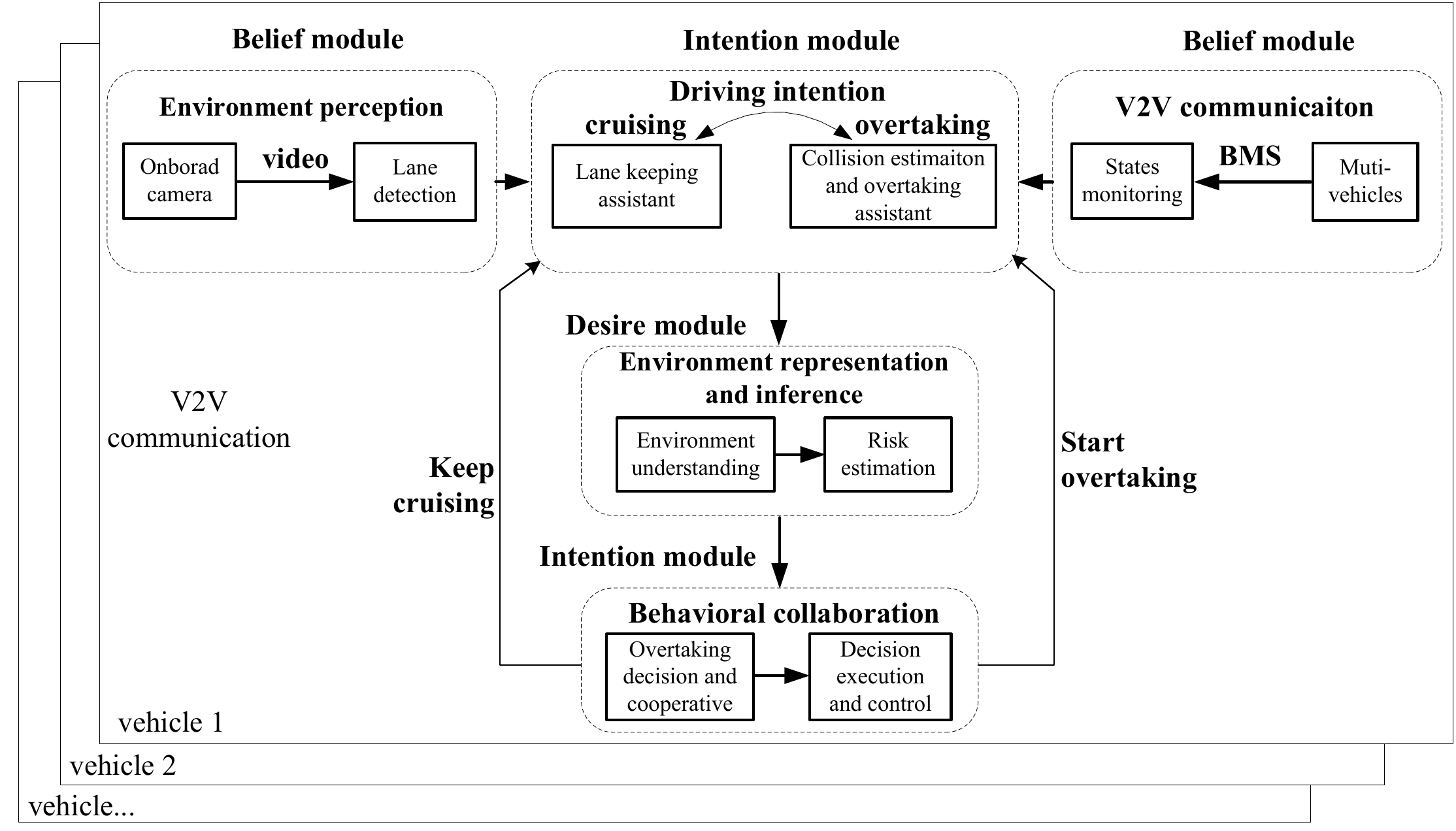}
{BDI based V2V collaboration framework. \label{fig3}}

\section{Bezier based lane detection and modeling}
Lane detection and modeling is the key component of belief module, which can give us the representation of the nearby environment. However, there are lots of interference in the lane detection task, such as light unevenness, shadows of vehicles and buildings, water and stains, wears and tears, and belts on the roads. These obstructions yield great difficulties in the task of understanding lane markings \cite{RN4982}. Here, for the purpose of detecting and modeling the front lanes, both the Bezier curve and hybrid Gaussian anisotropic filter were adopted to increase the accurateness and robustness of the proposed method. With the Bezier curve, the different grades of roads can be flexibly modeled with corresponding degree of control points. Meanwhile, considering the anisotropy needs of road image preprocessing, along the road direction, we need a smoothing filter to eliminate image defects such as breakage, contamination, etc. and in the vertical direction, we need edge enhancement filters to enhance the characteristics of the road for subsequent road detection modules. 
\subsection{Image preprocessing and filtering}
In order to improve the robustness, two layers of ROI (region of interests) was set to avoid the noise interference in non-road areas and to improve the algorithm's real-time performance, which was shown in Figure 4. In the high-level layer, static ROI was set in the original image
\begin{equation}
R=\left( k\times \text{ImW},l\times \text{ImH},vpx+\Delta x,vpy+\Delta y \right)
\end{equation}

Where, ImgW and ImgH are the width and height of the image, and $k$, $l$ are proportional adjustment coefficients, ($vpx+\Delta x$ , $vpy+\Delta y$) is the coordinates of the center of the area of interest, $\Delta x$ , $\Delta y$ Is the deviation adjustment coefficient. 

When it comes to the low level, dynamic ROI was set on the bird's-eye view image according to the current vehicle status and driving intention, as shown in Figure 4. As soon as the lane changing behavior is detected from steering signal, the width of dynamic ROI W will be increased and the deviation coefficient a will be decreased, for the purpose of extending the search area of nearby lanes. On the other hand, the height of dynamic ROI H is determined by the current vehicle speed. When the speed is high, the speed coefficient b and the regional height H will be increased dynamically to enlarge the perception area in front of ego vehicle.

When it comes to the low level, dynamic ROI was set on the bird's-eye view image according to the current vehicle status and driving intention, as shown in Figure 4.  For the purpose of extending the search area of nearby lanes, as soon as the lane changing behavior is detected from steering signal, the width of dynamic ROI W will be increased and the deviation coefficient $a$ will be decreased. Meanwhile, the height of dynamic ROI H is determined by the current vehicle speed. When the speed is high, the speed coefficient $b$ and the regional height H will be increased dynamically to enlarge the perception area in front of ego vehicle.

In the above image transformation step, according to the assumption of the pinhole camera model, the transformation matrix from vehicle to the camera can be calculated by the following equation.

\begin{equation}
T=\left[ \begin{matrix}
   -\frac{h}{{{f}_{u}}}{{c}_{2}} & \frac{h}{{{f}_{v}}}{{s}_{\text{1}}}{{\text{s}}_{2}} & \frac{h}{{{f}_{u}}}{{c}_{u}}{{c}_{2}}-\frac{h}{{{f}_{v}}}{{c}_{v}}{{s}_{1}}{{s}_{2}}-h{{c}_{1}}{{s}_{2}} & 0  \\
   \frac{h}{{{f}_{u}}}{{s}_{2}} & \frac{h}{{{f}_{v}}}{{s}_{1}}{{c}_{1}} & -\frac{h}{{{f}_{u}}}{{c}_{u}}{{c}_{2}}-\frac{h}{{{f}_{v}}}{{c}_{v}}{{s}_{1}}{{c}_{2}}-h{{c}_{1}}{{c}_{2}} & 0  \\
   0 & \frac{h}{{{f}_{v}}}{{c}_{1}} & -\frac{h}{{{f}_{v}}}{{c}_{v}}{{c}_{1}}+h{{s}_{1}} & 0  \\
   0 & -\frac{1}{{{f}_{v}}}{{c}_{1}} & \frac{1}{{{f}_{v}}}{{c}_{v}}{{c}_{1}}-{{s}_{1}} & 0  \\
\end{matrix} \right]
\end{equation}

Where, $T$ is the transformation matrix from vehicle coordinate to camera coordinate, the $f{{l}_{x}}$and$f{{l}_{y}}$ are the horizontal and vertical shift of camera focus; $o{{c}_{x}}$ and $o{{c}_{y}}$ are the horizontal and vertical position deviation of camera optical center; and ${{c}_{1}}=\cos ({{\theta }_{pitch}})$ ${{c}_{2}}=\cos ({{\theta }_{yaw}})$ ${{s}_{1}}=\sin ({{\theta }_{pitch}})$ ${{s}_{2}}=\sin ({{\theta }_{yaw}})$. ${{\theta }_{pitch}}$ and ${{\theta }_{yaw}}$ are defined in Figure 1. Using transformation matrix $T$, the inverse perspective transformation image can be easily obtained. 

Hybrid Gaussian anisotropic filter was also used for improving the robustness of the proposed method in the image preprocessing step. Taking Gaussian function as scale function, a hybrid Gaussian anisotropic filter was constructed by using low-pass smooth Gaussian filter and second-order Mexican hat wavelet high-pass filter, as the following equation:
\begin{equation}
{{\text{G}}^{\theta }}\text{=}G_{x}^{{{0}^{\circ }}}\cos \theta +G_{y}^{{{90}^{\circ }}}\sin \theta 
\end{equation}

Where,        
\begin{equation}
G_{v}^{{{90}^{\circ }}}=\exp (-\frac{1}{2\sigma _{v}^{2}}{{v}^{2}})                                   
\end{equation}
\begin{equation}
G_{u}^{{{0}^{\circ }}}=\frac{1}{\sigma _{u}^{2}}(1-\frac{{{u}^{2}}}{\sigma _{u}^{2}})\exp (-\frac{1}{2\sigma _{u}^{2}}{{u}^{2}})
\end{equation}

In the above equations, equation 4 is a Gaussian Lowpass Filter and equation 5 is a second order Mexican hat wavelet high pass filter. $\theta $ is the angle input of filter direction, $\sigma _{u}^{2}$ depends on the width of the front lane, $\sigma _{v}^{2}$ depends on the length of road on the dynamic ROI.

\Figure(topskip=5pt, botskip=0pt, midskip=3pt)[width=242pt]{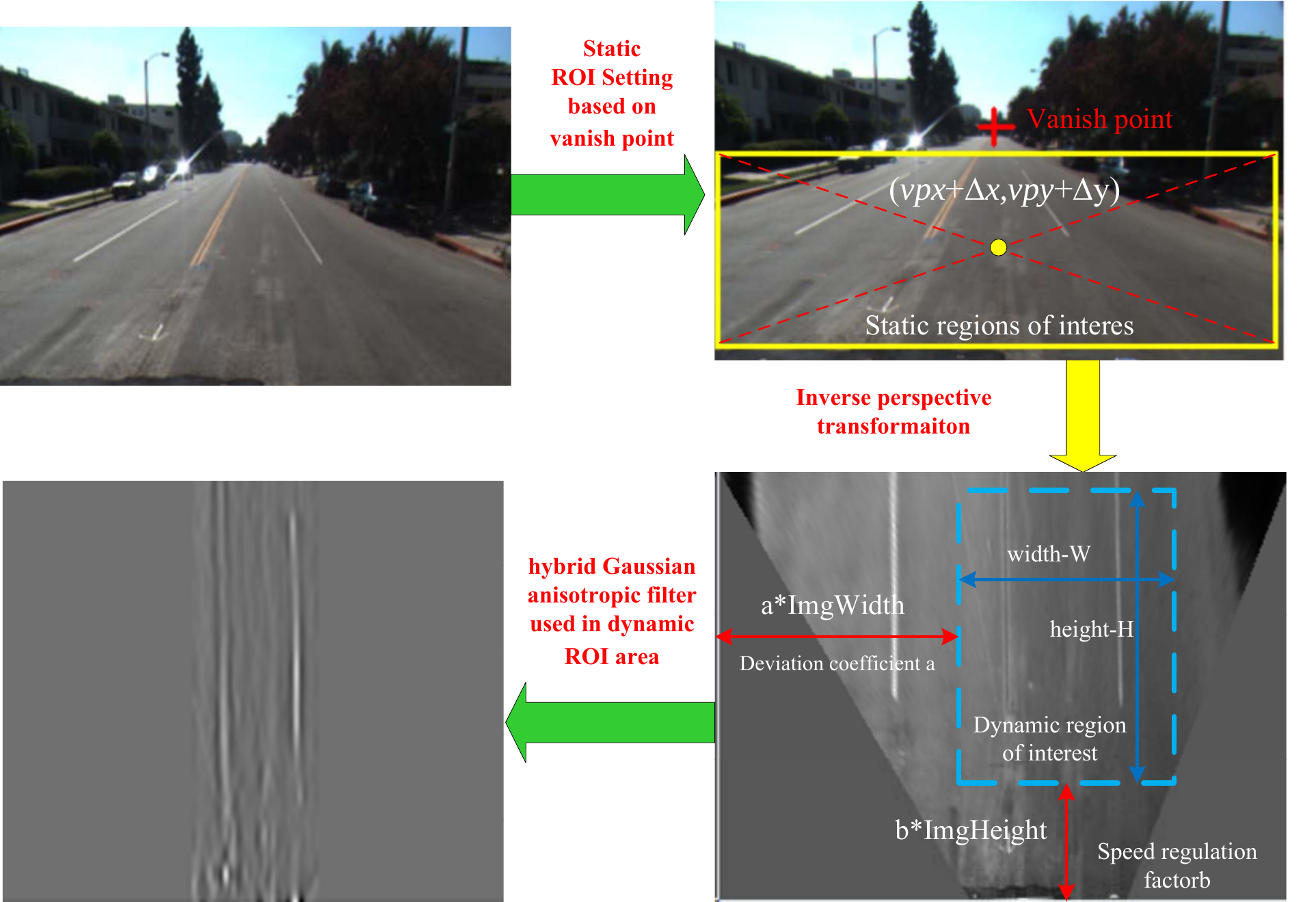}
{Image preprocessing and the setting of two layer ROI. \label{fig4}}

\subsection{Bezier based uncertain deformation template}
Commonly, driving roads can be modeled by different curve types, such as straight line model, quadratic curve model and higher order curve model\cite{RN5831}. While, simple line model was widely used in the highway area, but it is hard to fit complex road. The quadratic curve has a unidirectional boundary curvature resulting in poor model adaptability. Higher order curve model can be used to successfully describe the complex road, but it has an unbearable computational complexity. Therefore, in this paper, the Bezier spline curve was adopted to construct the Uncertain Deformation Template (UDT) for complex lane detection, which can automatically choose the complexity of model types. 

The Bezier curve is constructed by Bernstein basis function, and the characteristics of the curve can only be determined by the position of its control points \cite{RN5836}. The definition of the Bezier curve is shown in the following equation: 
\begin{equation}
P(t)=\sum\limits_{i=0}^{n}{{{P}_{i}}\frac{n!}{i!\left( n-i \right)!}{{\left( 1-t \right)}^{n-i}}{{t}^{i}}}\begin{matrix}
   {} & (0\le t\le 1)  \\ \end{matrix}
\end{equation}

From the definition, we can easily found that the different degree of Bezier equations can be modeled with different n values. In detail, commonly, Linear Bezier curve $P\left( t \right)=\left( 1-t \right){{P}_{0}}+t{{P}_{1}}$ can be used to raise the straight road type for highway, and Quadratic Bezier curve $P\left( t \right)={{\left( 1-t \right)}^{2}}{{P}_{0}}+2t\left( 1-t \right){{P}_{1}}+{{t}^{2}}{{P}_{2}}$, which is constructed by two control points can be used to build the curve road. Cubic Bezier curve $P\left( t \right)={{\left( 1-t \right)}^{3}}{{P}_{0}}+3t{{\left( 1-t \right)}^{2}}{{P}_{1}}+3{{t}^{2}}\left( 1-t \right){{P}_{2}}+{{t}^{3}}{{P}_{3}}$ can be used to put up S-shape path. 

Thereafter, using the different degree of Bezier spline curve, a UDT can be constructed, which was shown in equation 7. Then, the land detection question can be transformed to the question of determining the parameters of UDT. 

\begin{equation}
L=\left[ n,{{P}_{n}},c,s \right]\begin{matrix}
   {} & (2\le n  \\
\end{matrix}\le 4)
\end{equation}

Where, $n$ is the order of the Bezier curve, ${{P}_{n}}$ is the curve control points determined by order, $c$ is the color of lane marking, $s$ is the credibility evaluation coefficient of the current template, and is normalized to the interval of [0 1].

\Figure(topskip=0pt, botskip=0pt, midskip=0pt)[width=242pt]{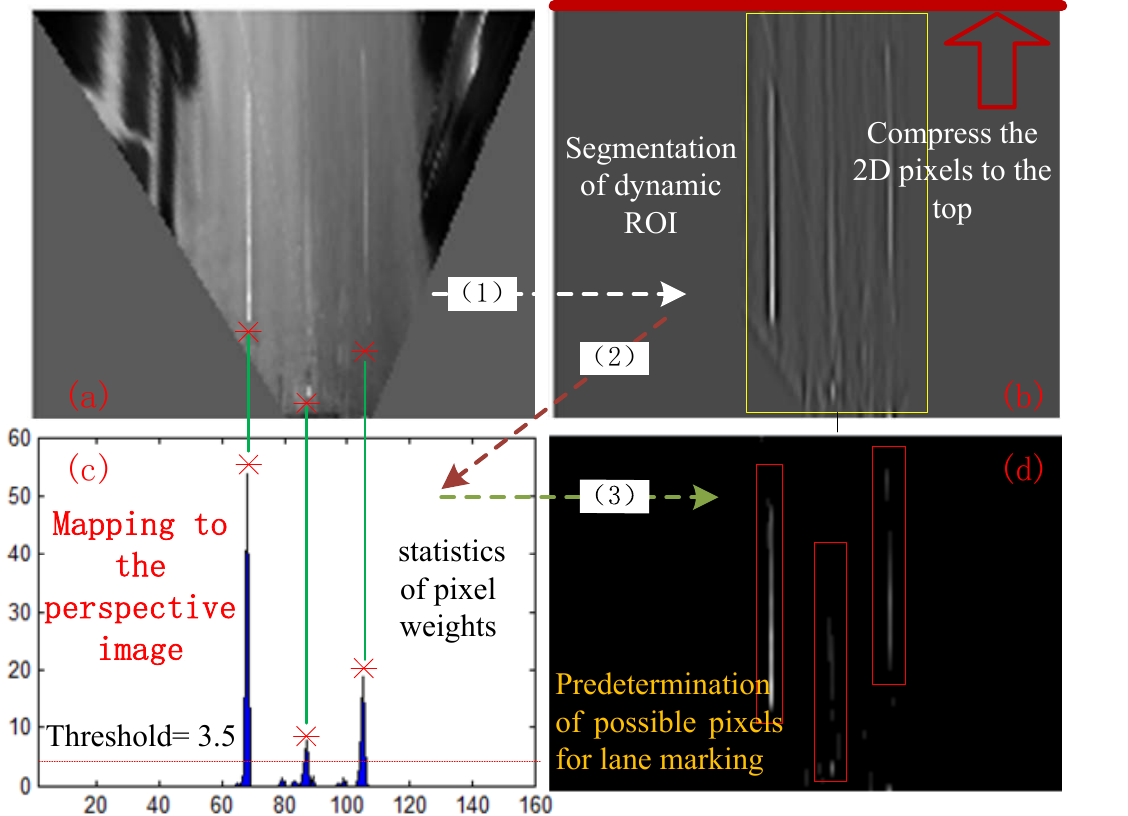}
{Data space setting of possible pixel belonging to lane marking. \label{fig5}}

\subsection{Parameters solving of UDT under hypothesis and testing problem}
In order to solve the hypothesis and testing problem, some of the previous works consider that every point in the image has the opportunity of belonging to the lane marking, so they make all the image points involved in the calculation of posterior probability density function, which consumes a lot of computing resources. \cite{RN5834}. In order to reduce the calculation, the data set of possible lane marking pixels was built, which was shown in Figure 5 with three steps. First, the dynamic ROI area was segmented from the original image for the further process, including hybrid Gaussian anisotropic filtering. Then, we compressed the 2D pixels in the processed image ROI into the image top and sum up the intensity of pixels in each image columns. Further, in the 1D intensity image, a reasonable threshold was set to predetermine the possible pixels belonging to the lane marking. In this way, the data space of possible lane marking pixels can be obtained, as shown in the following equation.
\begin{equation}
\Omega =\left\{ {{P}_{1}},{{P}_{2}},\cdots {{P}_{M}} \right\}
\end{equation}

Where, $\Omega $ is the sample space, $M$ is the number of samples, ${{P}_{i}}$ is the possible lane marking pixel. The possible pixel picking method can effectively avoid the negative impact of many road noises, such as light unevenness, road wears, and tears, etc. 

In the hypothesis step, the Random Sample Consensus (RANSAC) algorithm was adopted to obtain the parameter hypothesis of current UDT. Thinking of the driving environment, most of roads are straight types, followed by curves, and the complicated road types are few. So, the template parameters are estimated by increasing the template order parameter $n$ from 2 to 4, which means that the parameter estimation process follows the logic from simple to complex. In details, first, N samples are randomly selected from the sample space to form a sample group, and this operation needs to be repeated Q times to obtain Q sample groups. By fitting the sample points in each sample set, Q fitting curves can be obtained. Then, the reliability of the Q fitting results are verified separately. If anyone of the Q fitting curves passes the consistency test, the search stops, and the curves with the highest credibility would be considered as the road fitting curve. Otherwise, N takes N+1 to repeat the above operation. Where the value of N directly determines the number of deformation template levels. It should be pointed out that when N is 2, the sample is fitted with a line; When N is 3, the parabola is used to fit the sample; The least squares fitting sample is used when N is 4. 

During the process, the mathematical expression of the Bezier curve can be written in the matrix form as shown below:

\begin{equation}
\begin{aligned}
  & {{\left[ \begin{matrix}
   {{Q}_{1}}  \\
   \cdots   \\
   {{Q}_{n}}  \\
\end{matrix} \right]}_{(n+1)\times 1}}= \\ 
 & {{\left[ \begin{matrix}
   t_{1}^{n} & \cdots  & 1  \\
   {} & \cdots  & {}  \\
   t_{n}^{n} & \cdots  & 1  \\
\end{matrix} \right]}_{(n+1)\times (n+1)}}{{M}_{(n+1)\times (n+1)}}{{\left[ \begin{matrix}
   {{P}_{1}}  \\
   \cdots   \\
   {{P}_{n}}  \\
\end{matrix} \right]}_{(n+1)\times 1}} \\ 
\end{aligned}
\end{equation}

It can be abbreviate represented as:
\begin{equation}
{{Q}_{n}}=TM{{P}_{n}}
\end{equation}

The solution to this equation is 

\begin{equation}
{{P}_{n}}={{(TM)}^{-1}}{{Q}_{n}}
\end{equation}

In the credibility testing step, pixel consistency and curve likelihood were used to determine whether the current fitting parameters of UDT can pass the consistency test. Firstly, the pixel consistency examining module was used to verify the grayscale weight of pixel points within the hypothesized curve. If the pixel consistency coefficient $L(S)$ in the following equation was below the set threshold, the current parameters of UDT would lose the verification process.

\begin{equation}
L(S)=c*\sum\limits_{i=0}^{i=s}{val}
\end{equation}

Where $c$ is the color compensation factor. When the road marking is white, $c=1$, and if it is yellow, $c=1.5$. $S$ is the number of positive pixels passing through the fitting curve, $val$ is the grayscale weight of pixel points.
Meanwhile, the curve likelihood coefficient would evaluate the length and bending degree of the fitting curve for the current UDT, which should not deviate from the normal range. The curve likelihood coefficient $Q(S)$ can be calculated using the following equation: 

\begin{equation}
Q(S)={{k}_{1}}\frac{l}{v}+\frac{{{k}_{2}}}{N-2}\sum\limits_{i=1}^{i=N-2}{\cos (\pi -{{\theta }_{i}})}
\end{equation}
   
Where, ${{k}_{1}}$is the length coefficient, $l$  is the distance between the furthest two sample points on the curve, $v$ is the image height, ${{k}_{2}}$ is the angle coefficient, $N$ is the number of sample points, ${{\theta }_{i}}$ is the angle between adjacent sample points. 

Then, use both of the pixel consistency and curve likelihood coefficients, we can construct a reliability evaluation index $s$ of UDT, which is shown in the following equation:

\begin{equation}
s={{k}_{L}}\times L(S)+{{k}_{Q}}\times Q(S)
\end{equation}

\Figure(topskip=0pt, botskip=0pt, midskip=0pt)[width=400pt]{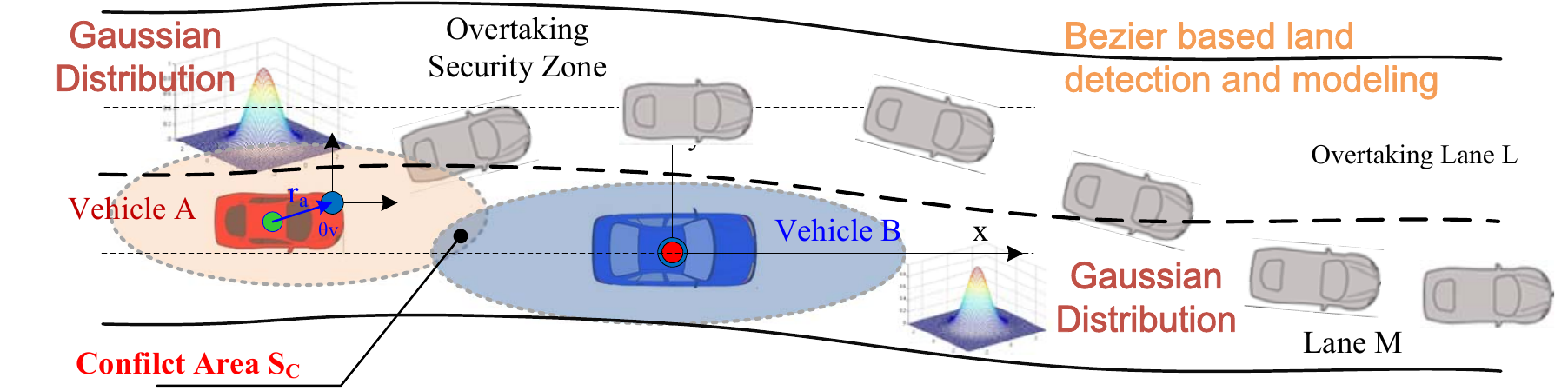}
{The brief introduction of overtaking procedure. \label{fig6}}

Here, ${{k}_{L}}$ and ${{k}_{Q}}$ are the weight proportionality coefficient of pixel consistency coefficient $L(S)$ and $Q(S)$ . In this way, we can reckon the parameters of UDT in real time using the RANSAC based hypothesis and testing method. 

\section{Gaussian-based conflict probability estimation}
With the models of the ahead lanes and the relative positions and velocities of nearby vehicles received from V2V sensors, the local dynamic environment can be built successfully. Using the time to collision (TTC) method, we can determine the proper time for the behavior decision of the overtaking. It should be noticed that the procedure should never be started if the front overtaking lane has been occupied by other slower vehicle or there is no enough space for the ego vehicle. Based on constant speed hypothesis, the TTC can be calculated by the following equation.

\begin{equation}
TTC=\frac{{{S}_{ab}}-{{L}_{a}}/2-{{L}_{b}}/2}{{{v}_{a}}-{{v}_{b}}}\ \ \ \ \text{(}{{v}_{a}}>{{v}_{b}}\text{)}  
\end{equation}

Where, ${{S}_{ab}}$ is the starting distance between ego vehicle A and nearby vehicle B. ${{L}_{a}}$ is the length of vehicle A and ${{L}_{b}}$ is the length of vehicle B. ${{v}_{a}}$ is the current speed of vehicle A and ${{v}_{b}}$ is the current speed of vehicle B.

During the process of overtaking, Gaussian-based conflict potential field was proposed to guarantee overtaking safety, which can be used to quantitatively estimate the oncoming collision danger. The brief introduction of overtaking procedure is shown in Figure 6.

\subsection{Conflict Potential Field}
During the process of overtaking, the uncertainty of collision risk can be represented using the probability density function, as shown in Figure 7. Conflict potential fields can be established respectively, while, the distributions of potential fields are assumed to satisfy multivariate Gaussian distribution with the direction pointing from the reference center to the opposite.

\Figure[h](topskip=0pt, botskip=0pt, midskip=0pt)[width=240pt]{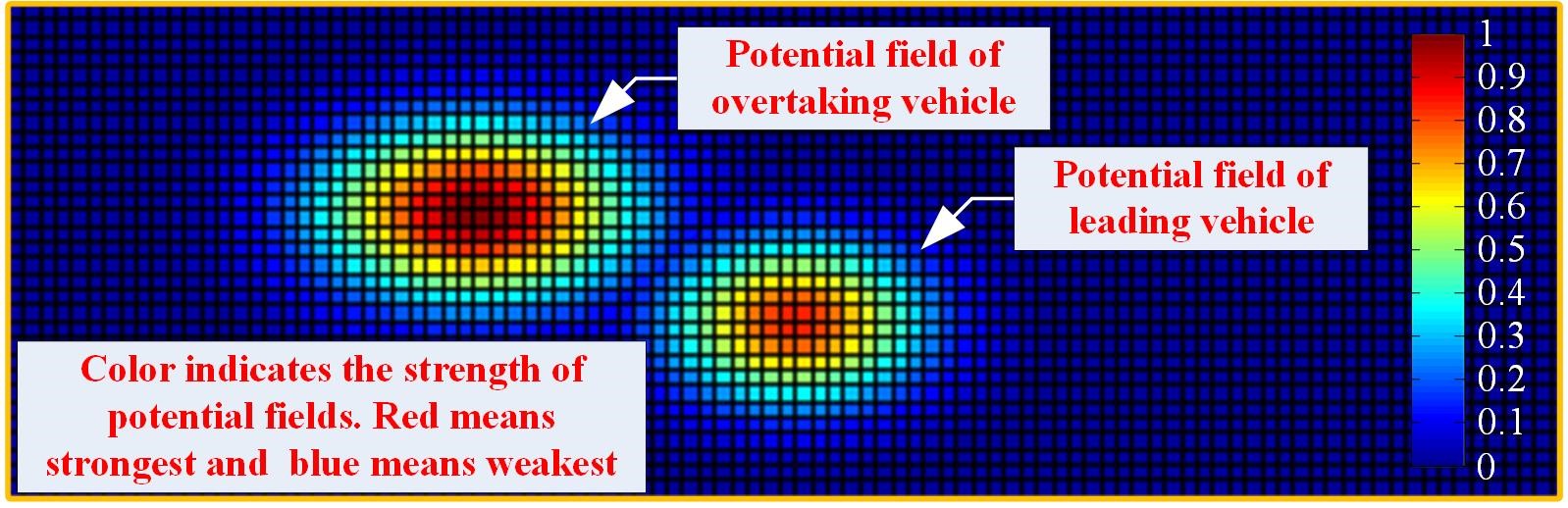}
{Potential fields at overtaking stage. \label{fig7}}

Taking overtaking vehicle (OV) A as an instance, the potential field along the major principal axis and minor principal axis are independent. The probability density function of the potential field of OV A can be denoted by a multivariate Gaussian distribution. As given by the following formula:
\begin{equation}
{{\mathbf{X}}_{A}}\left| {{\mathbf{\mu }}_{A}},{{\mathbf{\Lambda }}_{A}} \right.)=\frac{1}{{{(2\pi )}^{D/2}}{{\left| {{\mathbf{\Lambda }}_{A}} \right|}^{1/2}}}\exp (-\frac{1}{2}\Delta _{A}^{2})
\end{equation}

Where, $N({{\mathbf{X}}_{A}}\left| {{\mathbf{\mu }}_{A}},{{\mathbf{\Lambda }}_{A}} \right.)$ is the probability density distribution of the conflict potential field, ${{\mathbf{X}}_{A}}$ is the input of two-dimensional variable, ${{\mathbf{\Lambda }}_{A}}$ is covariance matrix, $\left| {{\mathbf{\Lambda }}_{A}} \right|$ is the determinant of ${{\mathbf{\Lambda }}_{A}}$, $D$ is the dimension value of input variables, in this paper $D=2$, ${{\mathbf{\mu }}_{A}}$ is the mean-variance of two-dimensional Gaussian distribution, ${{\Delta }_{A}}$ is the Mahalanobis distance from ${{\mathbf{\mu }}_{A}}$ to ${{\mathbf{X}}_{A}}$ and the calculation is given by:

\begin{equation}
{{\Delta }_{A}}^{2}={{({{\mathbf{X}}_{A}}-{{\mathbf{\mu }}_{A}})}^{\text{T}}}{{\mathbf{\Lambda }}_{A}}^{-1}({{\mathbf{X}}_{A}}-{{\mathbf{\mu }}_{A}})
\end{equation}

The distribution of potential fields in the major principal axis and minor principal axis are independent of each other. Then, the covariance matrix of the potential field can be obtained as:

\begin{equation}
{{\mathbf{\Lambda }}_{\mathbf{A}}}=\left[ \begin{matrix}
   \sigma _{Ax}^{2} & 0  \\
   0 & \sigma _{Ay}^{2}  \\
\end{matrix} \right]
\end{equation}

Taking into account the impact of the relative speed, obviously, it would affect the collision risk and make the potential fields more deformable. So, the standard deviation of the covariance matrix ${{\sigma }_{Ax}}$ is constructed by two parts: basic value ${{\sigma }_{x}}$ and compensate value which is constructed by the relative longitudinal velocity ${{\delta }_{v}}^{{}}$. The standard deviation is given by:

\begin{equation}
{{\sigma }_{Ax}}=\sigma _{x}^{{}}\pm {{r}_{a}}*{{\delta }_{v}}^{{}}
\end{equation}

Where, ${{r}_{a}}$ is the gain coefficient of forwarding direction variance. Considering the impact of  ${{\sigma }_{Ax}}$ on the standard deviation of the lateral covariance matrix ${{\sigma }_{Ay}}$, the lateral covariance matrix ${{\sigma }_{Ay}}$ can be reckoned by equation (13), while assuming the linear relationship between ${{\sigma }_{Ax}}$ and ${{\sigma }_{Ay}}$ .

\begin{equation}
\sigma _{Ay}^{{}}=\min ({{r}_{c}}*{{\sigma }_{Ax}},\bar{\sigma }_{y}^{{}})
\end{equation}

Where, ${{r}_{c}}$ is the gain coefficient of lateral direction variance. ${{\bar{\sigma }}_{y}}$ is the saturation value for limiting ${{\sigma }_{Ay}}$ in a reasonable range. Formula (18) and formula (19) obviously implies that the relatively high velocity will not only increase the possibility of the collision before exceeding the overtaken vehicle but also decrease the possibility of collision after the exceeding. Similarly, the conflict potential field of overtaken vehicles can also be constructed in accordance with the method mentioned above.

\subsection{Estimation of Conflict Probability}

Based on the constructed conflict potential fields at the surpassing stage, the estimation of conflict probability can be calculated by integrating the probability density of potential fields over the conflict area, which is shown in Figure 6.

In order to simplify the calculation, the two conflict potential fields of overtaking vehicle A and leading vehicle B can be uniformed to the world coordinate system. The transformation matrix from the vehicle coordinate system to the world coordinate system is given by:

\begin{equation}
R=\left[ \begin{matrix}
   \cos \theta  & -\sin \theta   \\
   \sin \theta  & \cos \theta   \\
\end{matrix} \right]
\end{equation}

Where, $R$ is the transformation matrix. $\theta $ is the azimuth angle between the vehicle coordinate and world coordinate. Therefore, the covariance matrix can be transformed from the vehicle coordinate to the world coordinate, as given by:

\begin{equation}
{{\mathbf{\Lambda }}_{\text{A}}}^{W}={{R}_{A}}{{\mathbf{\Lambda }}_{A}}{{R}_{A}}^{\text{T}}
\end{equation}
\begin{equation}
{{\mathbf{\Lambda }}_{B}}^{W}={{R}_{B}}{{\mathbf{\Lambda }}_{B}}{{R}_{B}}^{\text{T}}
\end{equation}

 Considering the irrelevant of the distribution of two potential fields, the joint error covariance matrix can be obtained according to the synthesis rules of multivariate Gaussian distribution, as follows:

\begin{equation}
\mathbf{\Lambda }={{\mathbf{\Lambda }}_{\mathbf{A}}}^{W}+{{\mathbf{\Lambda }}_{B}}^{W}
\end{equation}

In order to facilitate the integration of conflict area, a conflict coordinate is established by taking the center of the potential field of leading vehicle B as the new origin, the major principal axis of conflict ellipses as new abscissa and the minor principal axis as new ordinate. Then, the center offset of the potential field of  overtaking vehicle A  can be given by:

\begin{equation}
{{\mathbf{\mu }}_{A}}=\left[ \begin{matrix}
   {{x}_{r}}  \\
   {{y}_{r}}  \\
\end{matrix} \right]
\end{equation}

Where,${{x}_{r}}$is the center offset of  overtaking vehicle A  along the abscissa direction of conflict coordinate, ${{y}_{r}}$ is the center offset in the conflict coordinate. According to the characteristic of the multidimensional normal distribution, a linear combination of normal distribution still meets the normal distribution. Hence, the joint probability density function of conflict can be given by: 

\begin{equation}
f(x,y)={{(2\pi )}^{-D/2}}{{\left| \mathbf{\Lambda } \right|}^{-1/2}}\exp (-\frac{1}{2}{{(\mathbf{X}-\mathbf{\mu })}^{\text{T}}}{{\mathbf{\Lambda }}^{-1}}(\mathbf{X}-\mathbf{\mu }))
\end{equation}

Then, the probability of collision at time t can be obtained through the integration of conflict probability density over the conflict area:

\begin{equation}
{{\text{S}}_{cp}}=\iint\limits_{{{S}_{c}}}{f(x,y)}dxdy
\end{equation}

Where, ${{\text{S}}_{\text{CP}}}$is the estimation of conflict probability. $f(x,y)$is the conflict probability density function. ${{S}_{c}}$ is the conflict area.

\section{Experiment Evaluation}
Driver on-loop tests were adopted for the experiment evaluation. The testing platform is shown in Figure 8, which contains 4 main components: the host machine (DELL Precision T5600 workstation), the target machine (Ubuntu OS based real-time PC), driving simulation interface and the information collection-control interface, which contains the driving control module, hydraulic braking module, road feeling module and warning module and collection control interface. With the responsibility of building multiple kinds of simulation environments and algorithms, the host machine uses a 3.9G Hz CPU, 32G RAM and a 2T ROM to guarantee the efficiency. 

\Figure[h](topskip=0pt, botskip=0pt, midskip=0pt)[width=240pt]{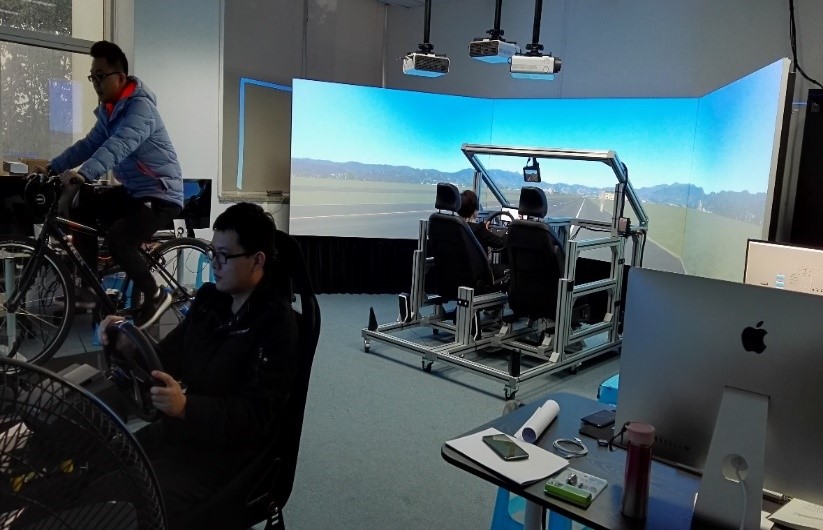}
{The platform of the cooperative collision simulation system. \label{fig8}}

In the test, we assumed that the vehicle dimension is of 1.8*4.2m (dwitdth *dlenth) and the maximum acceleration is of $\pm 2.7\ m/{{s}^{2}}$, and the maximum speed (vmax) is of 35m/s. The V2V sensors are installed to evaluate the relative position and distance between nearby vehicles. The effective information  transmission distance is about 200 m with the typical interval frequency of 10 Hz. The lane width is set to 3.5m. The communication mechanism conforms to the standard of SAE J2735 protocol.

\subsection{Bezier based lane detection and modeling}
\Figure[h](topskip=0pt, botskip=0pt, midskip=0pt)[width=240pt]{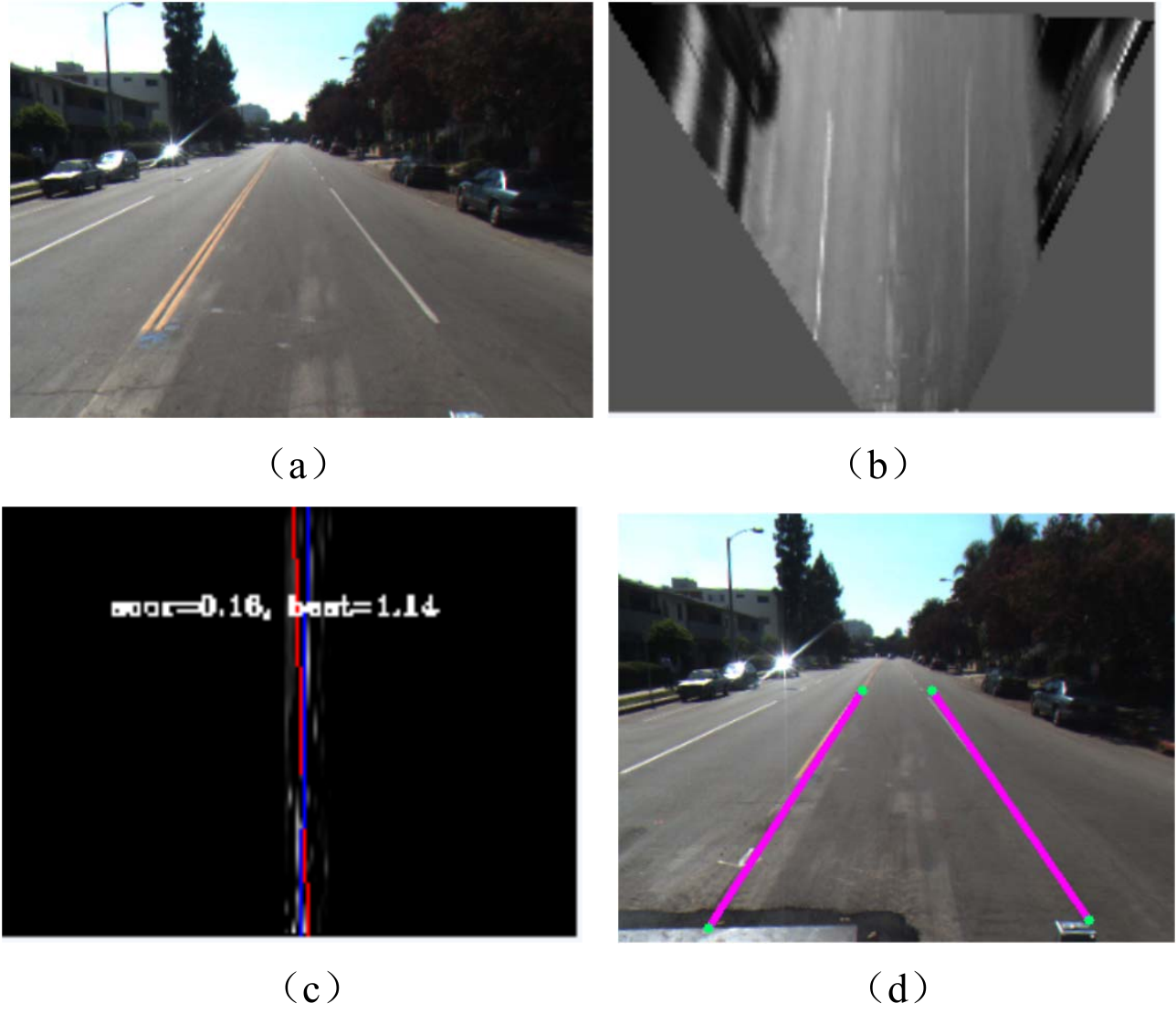}
{The recognition process of the road images. \label{fig9}}

In this testing part, 5932 frames of typical road pictures were tested for the proposed lane detection module and the results were statistically analyzed. The results show that the average recognition error rate of the algorithm is less than 6\%, the average detection time of each frame lane line is 50ms meeting the real-time requirements

Figure 9 graphically demonstrated the recognition process of the road images. Where, figure 9(a) is the raw RGB image taking from onboard forward facing camera, and the time is in the evening with a good light condition. Figure 9(b) is the result of the image inverse perspective transformation. Fig.9 (c) is the RANSAC based parameters solving procedure of UDT under hypothesis and testing problem. where the red line is the curve fitted by the currently randomly selected deformation template. At this time, N=2, the current template matching credibility is 0.16, and the blue line is the best historical matching result in this search process, and the template matching degree is 1.14. Figure 9(d) is the final lane fitting model, and it has been transformed from the inverse perspective image to the original image.

\subsection{ Performance Evaluation of Cooperative Overtaking}
\subsubsection{Effects of the Conflict Area}
To look into the influence of the conflict area on the collision probability estimation, using the same overtaking sceneries, different scope of conflict areas were tested in the experiment, where the expected velocity of the leading vehicle is 55 km/h and that of the overtaking vehicle is 80 km/h. Figure 10 presents the results of collision probability estimation and the consuming time with different conflict areas, including ${{S}_{c}}$ = 17.5 m × 4.2 m, ${{S}_{c}}$ = 18 m × 4.2 m, ${{S}_{c}}$ = 18.5 m × 4.2 m, and ${{S}_{c}}$ = 19 m × 4.2 m.

As shown in Figure 10, the result shows that the proposed collision probability estimation value has not remarkably increased with the increase of the conflict areas. Moreover, the risk level of overtaking maneuver was not significantly affected by the variation of the conflict area and was kept in a certain scope.

\Figure[h](topskip=0pt, botskip=0pt, midskip=0pt)[width=242pt]{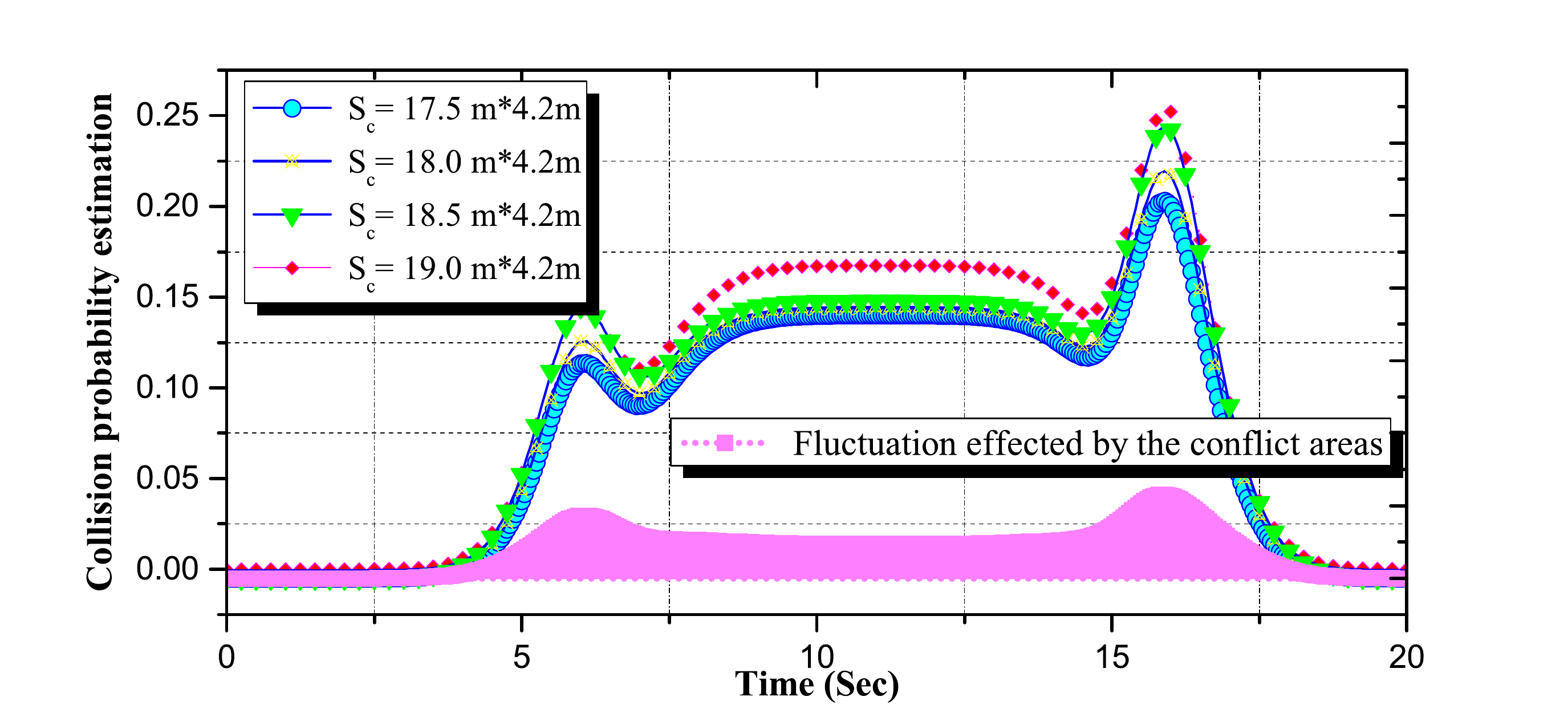}
{The effect of the conflict area to the collision probability. \label{fig10}}

\subsubsection{Effects of the Collision Types}
To investigate the effect of the collision types on the estimation of collision probability, different collision types have been tested in the experiment, where the relative collision velocity was set to 20km/h and the conflict areas was set to 17m×4 m. In the test, three typical collision modes including Rear-Rear collision (RRC), side-by-side collision (SRC), Front-to-Rear collision (FRC) were tested, as well as the normal overtaking maneuver for the comparison. 
The result in Figure11 shows that the proposed method has great adaptability to different collision types. Facing the inevitable oncoming collision, under different collision modes, the estimation curve of collision probability have a similar shape and trend, which is significantly useful for the collision avoidance technology and autonomous driving system. Obviously, if we set the overtaking prevention threshold to 0.3, then, more than 3s can be saved before the collision happens. If it is fully utilized, the collision accident would be greatly reduced.

\Figure[h](topskip=0pt, botskip=0pt, midskip=0pt)[width=242pt]{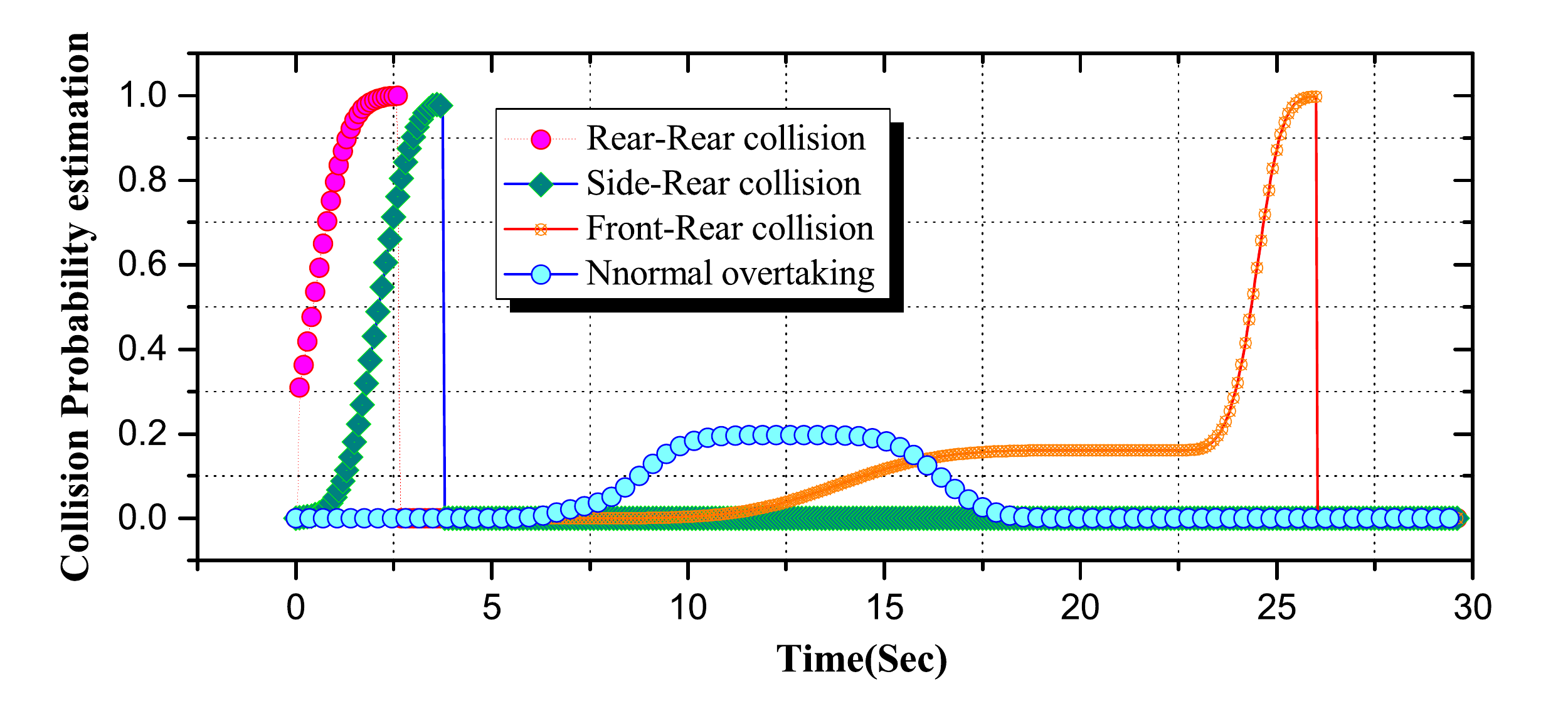}
{Collision probability estimation of different collision modes. \label{fig11}}

\subsection{ Case Evaluation and Discussion}

 Case 1: Straight Road with vehicles moving at different relative distances
 \Figure[h](topskip=0pt, botskip=0pt, midskip=0pt)[width=350pt]{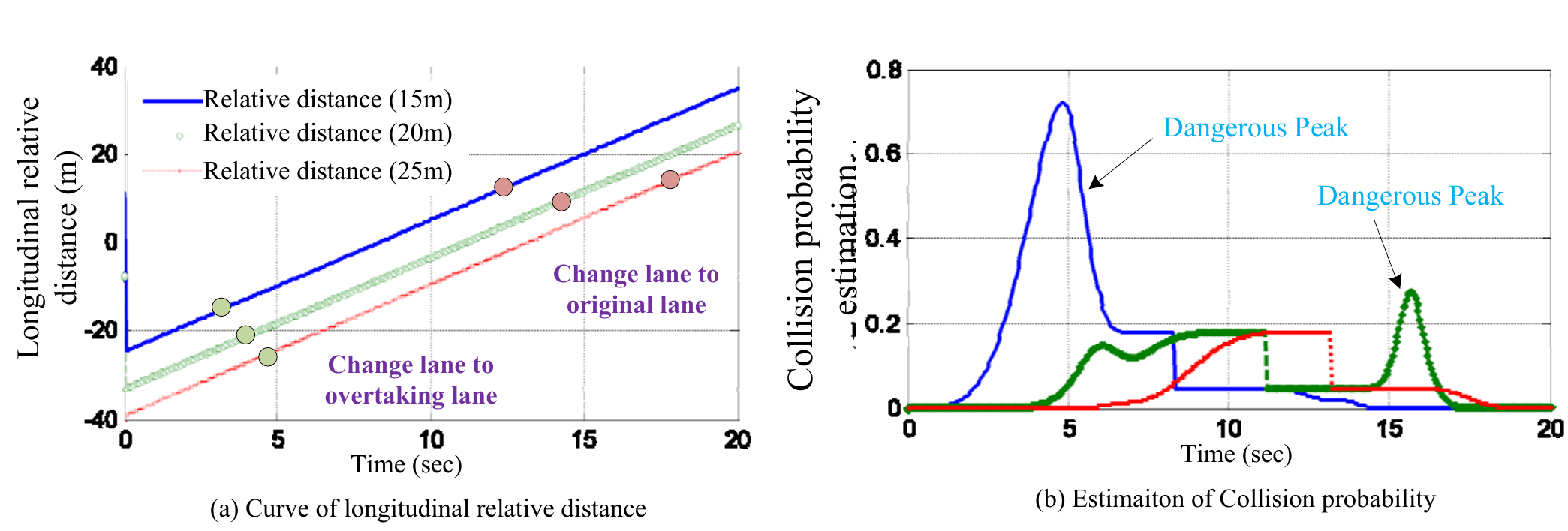}
{ Collision probability estimation with different relative distance. \label{fig12}}
\Figure[h](topskip=0pt, botskip=0pt, midskip=0pt)[width=350pt]{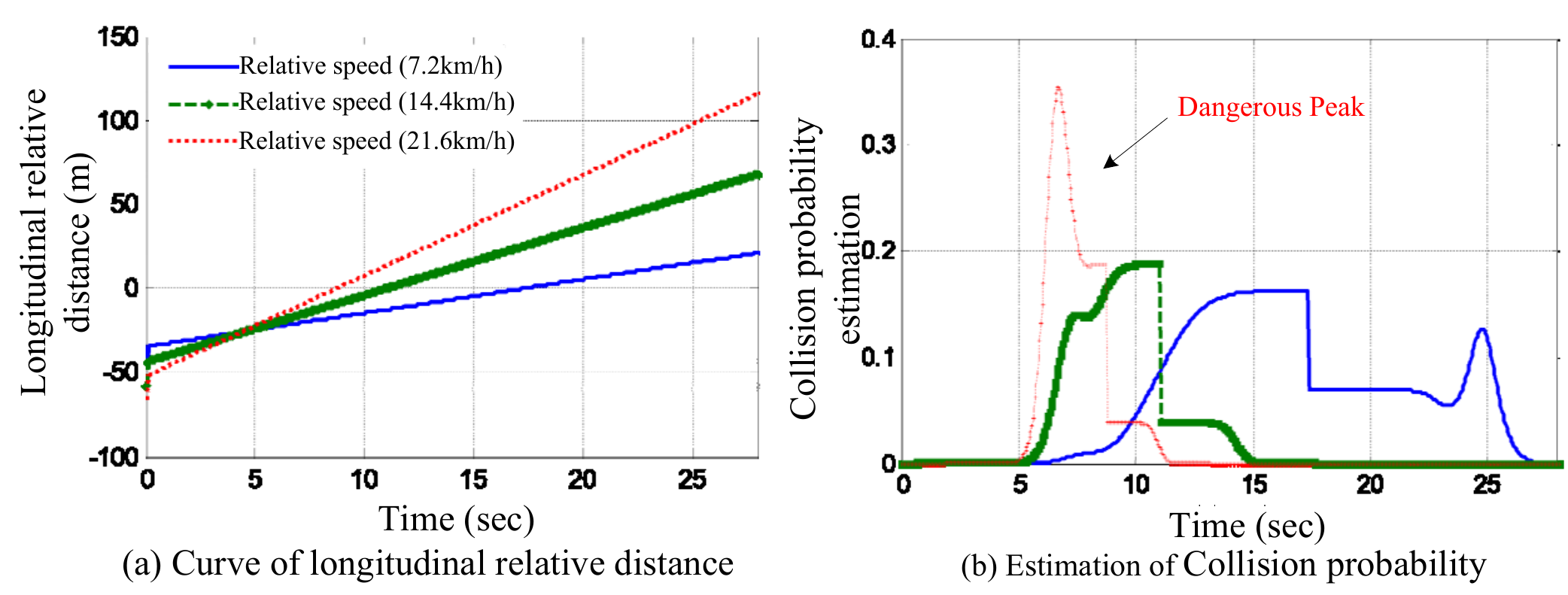}
{ Collision probability estimation with different relative speed. \label{fig13}}

In this case, a test environment of rural straight roads with medium concentration of fog was established. The visibility was only 120 m. The ego vehicle traveled with a slow speed and the maximum speed was just 35 km/h. Three straight lanes and four vehicles were set, including vehicle A (ego vehicle), vehicle B (forward leading vehicle), vehicle D (oncoming vehicle) and vehicle E (same direction vehicle). All of them traveled at constant speeds of 25, 18, 28 and 22 km/h respectively. Meanwhile, some slippery ice was added onto the surface of the road to decrease the adhesion coefficient of pavement. The visual-driver started up the overtaking action with different relative distances corresponding to the leading vehicles, which included 10m, 20m, and 25m. The aim of this test was to verify the capability of the proposed method to get correct and stabilized estimation of collision probability during the overtaking process. The test result was shown in Figure 12.

In the test, the ego vehicle started up the overtaking action at t$\sim$2s (before reminding signal), t$\sim$3s (after reminding signal and before warning signal) and t$\sim$5s (after warning signal) respectively, and returned back to the original lane after surpassing the leading vehicle. The simulation result shows that it would be very safe if the driver started the overtaking action before reminding signal, while the occupation time in the rapid lane would increase. On the other hand, as shown in Figure 12, if the driver started the overtaking action after warning signal, the collision risk would increase rapidly to 0.75, that portended a possible traffic accident would happen. If the driver complied with the instruction of the proposed method and start up the overtaking action at the right time, the collision risk would keep below 0.3 to guarantee the safety of the whole overtaking process.

Case 2: Curve Road with other vehicles moving at variable relative high speed on a sunshine day

In this case, a curved road was established with a minimum curvature radius of 150m. Meantime, the weather was changed to a good sunshine day and without lateral wind. In this weather and road condition, the influence of collision risk under variable relative speeds was tested to demonstrate the performance of the proposed method. According to the traffic rules, we set the maximum speed of the vehicles to 100 km/h and the minimum speed of 60 km/h, with the acceleration/deceleration of 3.8 m/s2. The aim of this simulation is to verify the performance of the proposed method under variable conditions of different relative speeds, which includes 7.2 km/h (low relative speed), 14.4 km/h (medium relative speed) and 21.6 km/h (high relative speed). The testing result was shown in Figure 13.

In the test, the ego vehicle cyclically detected the motion states of nearby vehicles in every 100ms, which was the same as the case 2, except that the communication distance was extended to 200m for the highway use. As shown in Figure 13, the ego vehicle started up the overtaking action at t$\sim$5s with three different relative speeds and it returned back to the original lane safely. The simulation result shows that, if the vehicle start the overtaking action with a low relative speed or a medium relative speed, collision risk would stay below the safety threshold of 0.3 to make sure the safety of the overtaking procedure. However, if the driver start up the overtaking action with a high relative speed of 21.6km/h, the collision risk would increase rapidly to 0.45, which is above the safety threshold. During the whole overtaking procedure, the proposed method can calculate the estimation of collision risk timely and correctly.

\section{Conclusions}
A novel methodology of cooperative overtaking based on BDI based multi-vehicle collaboration framework was proposed, which uses different kinds of heterogeneous sensors, to extend the awareness of the surrounding environment. The lane markings in front of ego vehicle were modeled with Bezier curves using the onboard cameras, which can adapt to different road types. While the nearby vehicles' position and velocity were obtained through the V2V communication scheme.  In addition, Gaussian-based conflict potential field was proposed to guarantee overtaking safety, which can be used to quantitatively estimate the oncoming collision danger. To support the proposed method, many experiments were conducted on the human-in-the-loop test. The results demonstrated that our proposed method achieves better performance, especially in some unpredictable nature road circumstances. In the future, we will focus on the implementation of the proposed method into the real vehicle and test its performance in the real road tests.

In addition I would like to know the predict performance in the real road test rather than the human-in-the-loop simulation test. 

\bibliography{your_external_BibTeX_file}

\begin{IEEEbiography}[{\includegraphics[width=1in,height=1.25in,clip,keepaspectratio]{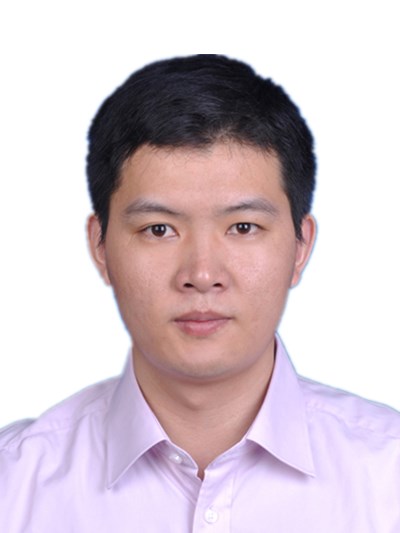}}] {Ke Wang} was born in Huaian, Jiangsu, China in 1984.  He received the B.S. and M.S. degrees in vehicle engineering from the Hunan University, Hunan, China, in 2007 and in 2009 and the Ph.D. degree in mechanical engineering from Hunan University, Hunan, China, in 2013 . 
From 2014 to 2016, he was an Assistant Professor with the Automobile Engineering Department. From 2016 to 2017, He finished his Postdoctoral research at College of Engineering, Michigan University Ann Arbor, USA. Since 2017, he has been an Associate Professor with the State Key Laboratory of Mechanical Transmission, Chongqing University. He is the author of one book, more than 20 articles, and more than 15 inventions. His research interests are the intelligent vehicle, environment perception and AI.
\end{IEEEbiography}
\begin{IEEEbiography}[{\includegraphics[width=1in,height=1.25in,clip,keepaspectratio]{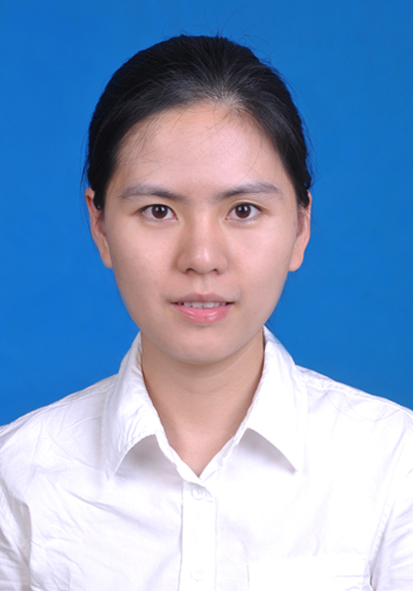}}]{Junlan Chen} was born in Zhuzhou, Hunan, China in 1985.She received the B.S. in Economics from the Northwestern Polytechnical University, Shanxi, China, in 2007 and she received her M.S. degrees and Ph.D. degree in Management from Hunan University, Hunan, China, in 2009 and 2013 respectively.
 From 2014 to 2016, she was an Assistant Professor with the School of Economics \& Management, Chongqing Normal University. From 2016 to 2017, she finished her Postdoctoral research at Research and Innovation Center, Ford motor company, Dearborn, USA. Until now, she is the author of more than 15 articles, and more than 10 inventions. Her research interests are the artificial intelligent, environment perception and economics in vehicle area.
\end{IEEEbiography}
\begin{IEEEbiography}[{\includegraphics[width=1in,height=1.25in,clip,keepaspectratio]{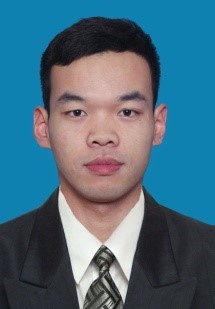}}]{Huanhuan Bao} was born in Qufu, Shandong, China in 1987. He received the B.S. degree in communication engineering from the Ludong University, Shandong, China, in 2011, and the M.S. degree in vehicle engineering from the Hunan University, Hunan, China, in 2014.
From 2015 to 2017, he was an engineer at the wind tunnel of China Automotive Engineering Research Institute. Since 2018, he was the head of the Department of Science and Technology. He is author of 3 articles and 5 inventions. His research interests are the intelligent vehicle, hydrogen fuel cell vehicle, and wind tunnel test.
\end{IEEEbiography}
\begin{IEEEbiography}[{\includegraphics[width=1in,height=1.25in,clip,keepaspectratio]{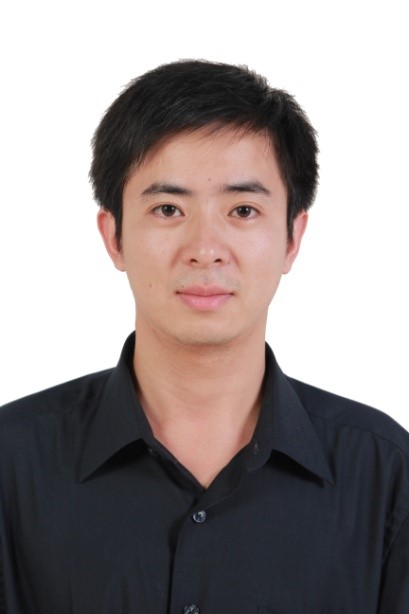}}]{Tao Chen} received the PhD. and Bachelor degree in Automotive Engineering from Tsinghua University. He is currently working as Deputy Director for Intelligent Vehicle Testing \& Evaluation Center, China Automotive Engineering Research Institute, and he is leading the V2X, Automated vehicle testing and engineering services to automotive industry. In his career from 2012 till now, he participated in several advanced research projects on Advanced Driver Assistant Systems(ADAS) , V2X, and automated vehicle development. He has published 20 SCI/EI indexed papers and owns 4 patents about intelligent vehicle technologies.
\end{IEEEbiography}
\EOD

\end{document}